\documentstyle[sprocl]{article}

\def\be{\begin{equation}}
\def\ee{\end{equation}}
\def\bea{\begin{eqnarray}}
\def\eea{\end{eqnarray}}

\def\L{{\cal L}}

\def\be{\begin{equation}}
\def\ee{\end{equation}}
\def\bea{\begin{eqnarray}}
\def\eea{\end{eqnarray}}

\begin{document}

\title{SELF-SIMILAR PERFECT FLUIDS}

\author{ J. CAROT and A.M. SINTES }

\address{ Departament de  F\'{\i}sica, Universitat de les Illes Balears,\\
 E-07071 Palma de Mallorca, Spain}


\maketitle\abstracts{Space-times 
admitting an $r$-parameter Lie group of homotheties are
 studied  for $r > 2$ devoting
 a special attention to those representing perfect fluid
solutions to Einstein's field equations.}



\section{Basic facts about homotheties}

Throughout this paper $(M,g)$ will denote a space-time: $M$ then being a
Hausdorff, simply connected, four-dimensional manifold, and $g$ a Lorentz
metric of signature (-,+,+,+). All the structures will be assumed smooth.

A global vector field $X$ on $M$ is called homothetic if either one of the
following conditions holds on a local chart
\be
\L_Xg_{ab}=2 \lambda g_{ab} \ ,\ \  \ X_{a;b}=  \lambda g_{ab}+F_{ab}\ ,
\label{r1}
\ee
where $\lambda$ is a constant on $M$ , $\L$ stands for the Lie derivative
operator, a semi-colon denotes a covariant derivative with respect to
the metric connection, and $F_{ab}=-F_{ba}$ is the so-called homothetic
bivector. If $\lambda \not=0$, $X$ is called proper homothetic and if
$\lambda=0$, $X$ is a Killing vector field (KV) on $M$. For a geometrical
interpretation of (\ref{r1}) we refer the reader to\cite{Hall88,Hall90b}.

A necessary condition that $X$ be homothetic is
 \be
{X^a}_{;bc}={R^a}_{bcd}X^d\ ,
\label{r2}
\ee
where ${R^a}_{bcd}$ are the components of the Riemann tensor in a
coordinate chart; thus, a homothetic vector field (HVF) is a particular
case of affine collineation\cite{HallCos88} and therefore it will satisfy
 \be
\L_X {R^a}_{bcd}=\L_X R_{ab}=\L_X{C^a}_{bcd}=0 \ ,\label{r3}
\ee
where $R_{ab}$ ($\equiv {R^c}_{acb}$) and ${C^a}_{bcd}$ stand, respectively
for the components of the Ricci and the conformal Weyl tensor.

The set of all  HVFs on $M$ forms a finite
dimensional Lie algebra under the usual bracket operation and will be
referred to as the homothetic algebra, ${\cal H}_r$, $r$ being its dimension.
The set of all Killing vectors fields on $M$ forms a finite dimensional Lie
algebra (dimension $s$) under the same bracket operation, and will be
referred to here as the Lie algebra of isometries, ${\cal G}_s$ which is
contained in (i.e., is a subalgebra of) ${\cal H}_r$. Furthermore, it is
immediate to see by direct computation that the Lie bracket of an HVF with a
KV is always a KV. From these considerations it immediately follows that the
highest
possible dimension of ${\cal H}_r$ in a four-dimensional manifold is $r=11
$.

If $r \neq s$ then $s=r-1$ necessarily, and one may choose a basis
$ \{X_A\}_{A=1\cdots r} \equiv  \{ X_1,\cdots, X_{r-1},X\} $
for ${\cal H}_r$,
in such a way that $X$ is proper homothetic and $X_1, \cdots,X_{r-1}$ are
Killing vector fields spanning ${\cal G}_{r-1}$. If these vector fields in
the basis of ${\cal H}_r$ are all complete vector fields, then each member
of ${\cal H}_r$ is complete and Palais' theorem
\cite{Palais57,Brickell70,Hall88b}
guarantees the existence of an $r$-dimensional Lie group of
homothetic transformations of $M$ ($H_r$) in a well-known way; otherwise,
it gives rise to a local group of local homothetic transformations of $M$
and, although the usual concepts of isotropy and orbits still hold, a little
more care is required \cite{Hall90}.

The following result \cite{Hall90,Bilyalov} will be useful: The orbits
associated with ${\cal H}_r$ and ${\cal G}_{r-1}$ can only coincide if
either they are four-dimensional or three-dimensional and null. (This
result still holds if ${\cal H}_r$ is replaced by the
conformal Lie algebra ${\cal C}_r$ and does not depend on the maximality of
${\cal H}_r$ or ${\cal C}_r$).

The set of zeroes of a proper HVF, i.e.,  $\{p\in M : X(p)=0\}$ (fixed points
of the homothety), either consists of topologically isolated points, or else
is part of a null geodesic. The latter case corresponds to the well-known
(conformally flat or Petrov type N) plane waves \cite{Hall88,Alex85}.

At any zero of a proper HVF on $M$ all Ricci and Weyl eigenvalues must
necessarily vanish and thus the Ricci tensor is either zero or has Segre
type $\{(2,11)\}$ or $\{ (3,1)\}$ (both with zero eigenvalue), whereas the
Weyl tensor is of the Petrov type $O$, $N$ or $III$ \cite{Hall88}
(see also \cite{Beem} for vacuum space-times).

\section{Basic facts about perfect fluids admitting HVFs}
The energy-momentum tensor for a perfect fluid is given by
\be
T_{ab}=(\mu+p)u_au_b+pg_{ab}\ ,
\label{r4}
\ee
where $\mu$ and $p$ are, respectively, the energy density and the pressure
as measured by an observer comoving with the fluid, and $u^a$ ($u^au_a=-1$)
is the four-velocity of the fluid. If $X$ is an HVF then, from Einstein's
Field Equations (EFE) it follows that
\be
\L_XT_{ab}=0\ ,
\label{r5}
\ee
and this implies in turn \cite{Eardley}
\be
\L_Xu_a=\lambda u_a\ , \quad \L_Xp=-2\lambda p\ , \quad \L_X\mu=-2\lambda
\mu\ . \label{r6}
\ee
Thus, the Lie derivatives of $u_a$, $p$ and $\mu$ with respect to a KV
vanish identically.

If a barotropic equation of state exists, $p=p(\mu)$,
and the space-time admits a proper HVF $X$ then \cite{Wainwright}
\be
p=(\gamma-1)\mu\ ,
\label{r7}
\ee
where $\gamma$ is a constant ($0\le\gamma\le 2$ in order to comply with
the weak and dominant energy conditions). Of particular interest are the
values $\gamma=1$
(pressure-free matter, \lq\lq dust") and $\gamma=4/3$ (radiation fluid).
In addition, the value $\gamma=2$ (stiff-matter) has been considered in
connection with the early Universe. Furthermore, values of $\gamma$
satisfying $0\le\gamma<2/3$, while physically unrealistic as regards a
classical fluid, are of interest in connection with inflationary models of
the Universe. In particular, the value $\gamma=0$, for which the fluid can
be interpreted as a positive cosmological constant, corresponds to
exponential inflation, while the values $0<\gamma<2/3$ correspond to power
law inflation in FRW models \cite{Barrow86}, but it is customary to
restrict $\gamma$ to the range $1\le\gamma\le 2$.

If the proper HVF $X$ and the four-velocity $u$ are mutually orthogonal
(i.e., $u^aX_a=0$) and a barotropic equation of state is assumed, it follows
that $\gamma=2$, i.e., $p=\mu$ stiff-matter \cite{Eardley}, on the other
hand, if $X^a=\alpha u^a$ the fluid is  then shear-free. Further information
on this topic can be found in \cite{McInt76,McInt78,McInt79}.

\section{The \lq\lq dimensional count-down"}
In this section, the maximal Lie algebra of global HVF on $M$ will be
denoted as ${\cal H}_r$ ($r$ being its dimension), and it will be assumed
that at least one member of it is proper homothetic.

The case of multiply transitive action is thoroughly studied in
\cite{Hall90}.
 We summarize in the following
table the results given there, which follow invariably from considerations on
the associated Killing subalgebra and the fixed point structure of the proper
HVF.

The first entry in the table gives the dimension of the group of
homotheties, the second and third entries stand for the nature and dimension
of the homothetic and Killing orbits respectively (e.g.: $N_2$, $T_2$ and
$S_2$ denote Null, Timelike and Spacelike two-dimensional orbits respectively,
$O_3$ stands for three-dimensional orbits of either nature, timelike,
spacelike or
null), the fourth and fifth entries give the Petrov and Segre type(s) of the
associated Weyl and Ricci tensors. Finally, the last two entries give
respectively the possible interpretation whenever it is in some sense
unique, and the existence or non-existence of perfect fluid solutions for
that particular case, along with some supplementary information; thus
FRW stands for Friedmann-Robertson-Walker, LRS for Locally Rotationally
Symmetric, and Bianchi refers to that family of perfect fluid solutions.
The cases that cannot arise are labeled as \lq\lq Not Possible", and
wherever no
information is given on the Petrov and Segre types, it is to be understood
that all types are possible in principle. The Segre type of the Ricci tensor
of the case described in the last row, is unrestricted except in that it
must necessarily have two equal (spacelike) eigenvalues; perfect fluid
solutions of these characteristics constitute special cases of spherically,
plane or hyperbolically symmetric perfect fluid space-times. For further
information
on LRS spacetimes, see \cite{Ellis67,Stewart68}; for the case $r=4$ transitive
and null three-dimensional Killing orbits, see \cite{Kramer,Rosq}. Regarding
spatially homogeneous Bianchi models, see \cite{Rosq,Ryan,Hsu1}; and for the
last three cases occurring in the table, see
\cite{Kramer,Barnes79,Goenner70}.

{\footnotesize
\begin{center}
\begin{tabular}{|r|c|c|c|c|c|c|} \hline
$r$ & $O_m$ & $K_n$ & {\it Petrov} & {\it Segre} &{\it Interpretation} &
{\it PF info.} \\ \hline

$11$ & $M$ & $M$ & $O$ & $0$ & Flat & $\not\exists $ \\
$10$ & $M$ & $M$ & - & - & Not Possible & $\not\exists $ \\
$9$  & $M$ & $M$ & - & - & Not Possible & $\not\exists $ \\
$8$  & $M$ & $M$ & $O$ & $\{(2,11)\}$ & Gen. P. wave & $\not\exists $ \\
$7$  & $M$ & $M$ & $N$ & $0,\{(2,11)\}$ & Gen. P. wave & $\not\exists $ \\
$7$  & $M$ & $T_3$ & $O$ & $\{(1,11)1\}$ & Tachyonic Fl. & $\not\exists $
\\
$7$  & $M$ & $N_3$ & - & - & Not Possible & $\not\exists $ \\
$7$  & $N_3$ & $N_3$ & $O$ & $\{(2,11)\}$ & Gen. P. wave & $\not\exists $
 \\
$7$  & $M$ & $S_3$ & $O$ & $\{1,(111)\}$ &Perfect Fluid & FRW \\
$6$  & $M$ & $M$ & - & - & Not Possible & $\not\exists $ \\
$6$  & $N_3$ & $N_3$ & $N$ & $\{(2,11)\}$ & Gen. P. wave & $\not\exists $
 \\
$5$  & $M$ & $M$ & - & - & Not Possible & $\not\exists $ \\
$5$  & $M$ & $N_3$ & - & - & - & $\not\exists $ \\
$5$  & $N_3$ & $N_3$ & - & - & Not Possible & $\not\exists $ \\
$5$  & $M$ & $T_3$ & $D,N,O$ & $\{1,1(11)\},\{2,(11)\}$  &  & LRS \\
$5$  & $M$ & $S_3$ & $D,O$ & $\{(1,1)11\},\{(2,1)1\}$  &  & LRS \\
$4$  & $M$ & $N_3$ & $II, III, D,N,O$ & $\{(1,1)(11)\},\{(2,11)\}$ & Plane
waves  & $\not\exists$ \\
$4$  & $N_3$ & $N_3$ & - & - & Not Possible & $\not\exists $ \\
$4$  & $M$ & $T_3$ &  &  &  & Bianchi \\

$4$  & $M$ & $S_3$ &  &  &  & Bianchi \\
$4$  & $O_3$ & $N_2$ & $N,O$  & $\{3,1 \}, \{2,11\},\{(1,1)11\}$ &  &
$\not\exists$ \\
$4$  & $O_3$ & $T_2$ & $D,O$  & $\{(1,1)11\}$ &  & $\not\exists$ \\

$4$  & $O_3$ & $S_2$ & $D,O$  & $\{ - (11) \}$  &  & $\exists$
\\         \hline
\end{tabular}
\end{center}
}

The case $r=3$ has an associated Killing subalgebra ${\cal G}_2$ and the
respective dimensions of their orbits are 3 and 2 (see for instance
\cite{Ali1,Uggla92,Mars,Hewitt91} and references cited therein). When the
Killing subalgebra has null orbits, the metric is of Kundt's class
\cite{Kundt61} and perfect fluids are excluded. If the Killing orbits are
timelike, the solutions can then be interpreted as special
cases of axisymmetric stationary space-times \cite{Kramer,Mars},
 and if they are spacelike
as special cases of inhomogeneous cosmological solutions or cylindrically
symmetric space-times. In both cases, perfect fluid solutions have
been found.

\end{document}